\pdfoutput=1
\documentclass[lettersize,journal]{IEEEtran}
\usepackage{amsmath,amsfonts}
\usepackage{algorithmic}
\usepackage{array}
\usepackage[caption=false,font=footnotesize,labelfont=rm,textfont=rm]{subfig}
\usepackage{textcomp}
\usepackage{stfloats}
\usepackage{url}
\usepackage{verbatim}
\usepackage{graphicx}
\usepackage{cite}
\def\BibTeX{{\rm B\kern-.05em{\sc i\kern-.025em b}\kern-.08em
    T\kern-.1667em\lower.7ex\hbox{E}\kern-.125emX}}
\usepackage{balance}
\usepackage{makecell}
\usepackage{amsmath}
\begin{document}
\title{Saliency-Aware Spatio-Temporal Artifact Detection for Compressed Video Quality Assessment}
\author{Liqun Lin, Yang Zheng, Weiling Chen, Chengdong Lan, Tiesong Zhao
\thanks{This work was mainly supported by the Natural Science Foundation of China under Grant 62171134 and Grant 61901119. It was also supported by the Natural Science Foundation of Fujian Province under Grant 2022J02015 and Grant 2022J05117. (Corresponding author: C. Lan.)
	
L. Lin and T. Zhao are with the Fujian Key Lab for Intelligent Processing and Wireless Transmission of Media Information, College of Physics and Information Engineering, Fuzhou University, Fuzhou, Fujian 350116, China, and also with Fujian Science \& Technology Innovation Laboratory for Opto-electronic Information of China (e-mails: \{lin\_liqun, t.zhao\}@fzu.edu.cn).

Y. Zheng, W. Chen, C. Lan are with the Fujian Key Lab for Intelligent Processing and Wireless Transmission of Media Information, College of Physics and Information Engineering, Fuzhou University, Fuzhou, Fujian 350116, China (e-mails: \{211127117, weiling.chen, lancd\}@fzu.edu.cn).}}

\markboth{Journal of \LaTeX\ Class Files,~Vol.~18, No.~9, September~2020}%
{Compressed Video Quality Index Based on Saliency-Aware Spatio-Temporal Artifacts Detection}

\maketitle

\begin{abstract}
Compressed videos often exhibit visually annoying artifacts, known as Perceivable Encoding Artifacts (PEAs), which dramatically degrade video visual quality. Subjective and objective measures capable of identifying and quantifying various types of PEAs are critical in improving visual quality. In this paper, we investigate the influence of four spatial PEAs (\emph{i.e.} blurring, blocking, bleeding, and ringing) and two temporal PEAs (\emph{i.e.} flickering and floating) on video quality. For spatial artifacts, we propose a visual saliency model with a low computational cost and higher consistency with human visual perception. In terms of temporal artifacts, self-attention based TimeSFormer is improved to detect temporal artifacts. Based on the six types of PEAs, a quality metric called Saliency-Aware Spatio-Temporal Artifacts Measurement (SSTAM) is proposed. Experimental results demonstrate that the proposed method outperforms state-of-the-art metrics. We believe that SSTAM will be beneficial for optimizing video coding techniques.
\end{abstract}

\begin{IEEEkeywords}
Video quality assessment, saliency detection, Perceivable Encoding Artifacts (PEAs), compression artifact.
\end{IEEEkeywords}
\begin{figure}[!h]
	\centering
	\subfloat[]{\includegraphics[width=1in]{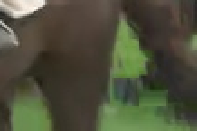}%
		\label{fig_11}}
	\hfil\vspace{-3mm}
	\subfloat[]{\includegraphics[width=1in]{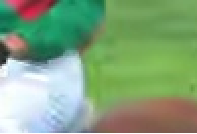}%
		\label{fig_12}}
	\hfil
	\subfloat[]{\includegraphics[width=1in]{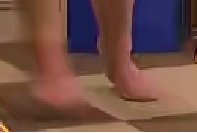}%
		\label{fig_13}}
	\hfil
	\subfloat[]{\includegraphics[width=1in]{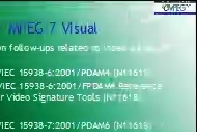}%
		\label{fig_14}}
	\hfil
	\subfloat[]{\includegraphics[width=1in]{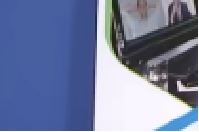}%
		\label{fig_15}}
	\hfil
	\subfloat[]{\includegraphics[width=1in]{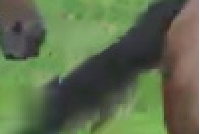}%
		\label{fig_16}}
	\caption{Examples of PEAs. (a) \pmb{Blocking:} Popular encoders are block-based, and all compression processes take place within non-overlapping blocks. This usually results in spurious discontinuities across block boundaries. The visual appearance of blocking may vary depending on the area of visual discontinuity \cite{3}. (b) \pmb{Blurring:} Video signal reconstruction results in a potentially significant loss of high frequency energy, which can lead to visual blurring. Perceptually, blurring usually manifests itself as a loss of spatial detail or sharpness of edge or texture areas in the image \cite{4}. (c) \pmb{Color bleeding:} It shows when the edges of one color in the video unintentionally overflow or overlap into another color \cite{5}. (d) \pmb{Ringing:} It appears in the form of ``halos", ``rings" or ``shadows" near sharp edges \cite{6}. (e) \pmb{Flickering:} It refers to frequent changes in brightness or chromaticity along the time dimension. It is very eye-catching and annoying to the viewer. (f) \pmb{Floating:} It is the illusory movement of certain areas while the surrounding areas remain stationary. Visually, the areas appear to float just above the surrounding background.}\par
	\label{fig1}
	\vspace{-0.3cm}
\end{figure}
\section{Introduction}
\IEEEPARstart{W}{ith} the growing demand of users, High-Definition (HD)/Ultra-HD (UHD) video is being used extensively. Due to the large amount of data, HD/UHD videos are encoded to meet the limited transmission bandwidth and storage space. The encoded videos have visually annoying Perceivable Encoding Artifacts (PEAs) which lead to visual quality degradation \cite{1}. Thus, Video Quality Assessment (VQA) based on PEAs is necessary for guiding the optimization of video coding to reduce distortion and improving the visual experience of users.

VQA can be divided into subjective and objective methods. For subjective VQA, Mean Opinion Score (MOS) and Different Mean Opinion Score (DMOS) suggested by the International Telecommunications Union (ITU) \cite{2} are used to reflect video subjective quality. Subjective quality assessment is accurate and reliable, but it is time-consuming and labor-intensive.

There are three types of objective VQA methods: Full Reference (FR), Reduced Reference (RR), and No Reference (NR) \cite{7}. The FR-VQA methods measure video quality by calculating the difference between the original videos and the compressed videos. Some FR methods like Peak Signal to Noise Ratio (PSNR) as well as Structural SIMilarity (SSIM) \cite{8} are widely used. The RR methods evaluate the quality of the compressed video by utilizing a few features of the original video. RR-VQA metrics like Spatial-Temporal RR Entropy Differences (STRRED) \cite{9} and Spatial Effective Entropy Difference Quality Assessment (SpEED-QA) \cite{10} also exhibited positive performance. However, these methods call for original video data that is typically absent from VQA, reducing the generalizability of models.

The most popular quality assessment scheme is NR-VQA because original videos are out of reach for end users. Some NR-VQAs such as Video Intrinsic Integrity and Distortion Evaluation Oracle (VIIDEO) \cite{11} and Two-Level Video Quality Model (TLVQM) \cite{12} improved the performance of video quality evaluation from different perspectives. Recently, more attention has been paid to the influence of compression artifacts on the compressed video quality. For instance, Bovik \cite{13} took into account video coding artifacts without specific classification. In \cite{6}, an NR-VQA method was developed by exploiting ringing and blocking artifacts. More than two types of spatial PEAs were considered in TLVQM \cite{12} and Saliency-Aware Artifact Measurement (SAAM) \cite{14}. The effects of blurring, blocking, and flickering artifacts on video quality are mentioned in \cite{15}, but no specific detection is performed.

As mentioned above, several VQA methods have taken into account compression artifacts, but most of them involve only a few types of compression artifacts. Moreover, these methods were very computationally intensive \cite{14}. To address these issues, we propose a method named Saliency-Aware Spatio-Temporal Artifacts Measurement (SSTAM). Here, we consider six typical PEAs which have a great impact on compressed video quality \cite{16}, as illustrated in Fig. \ref{fig1}. The main contributions are summarized as follows.

(1)An NR-VQA method is proposed based on PEA detection. The PEA detection module accurately recognizes six typical types of PEAs. Based on the PEA detection module, the PEA intensities are obtained to predict video quality.

(2)A saliency model that leverages visual saliency to achieve highly consistent with human visual perception is put forward. It makes our method more relevant to quality evaluation of compressed videos, while excluding regions that have very little impact on the video visual quality to reduce computational consumption.

(3)A comprehensive validation on four publicly available databases is conducted to verify our proposed SSTAM. The results reveal the promising performance of SSTAM in video quality evaluation.

\section{PEA-Based Video Quality Index}
\begin{figure*}[!t]
	\centering
	\includegraphics[width=7in]{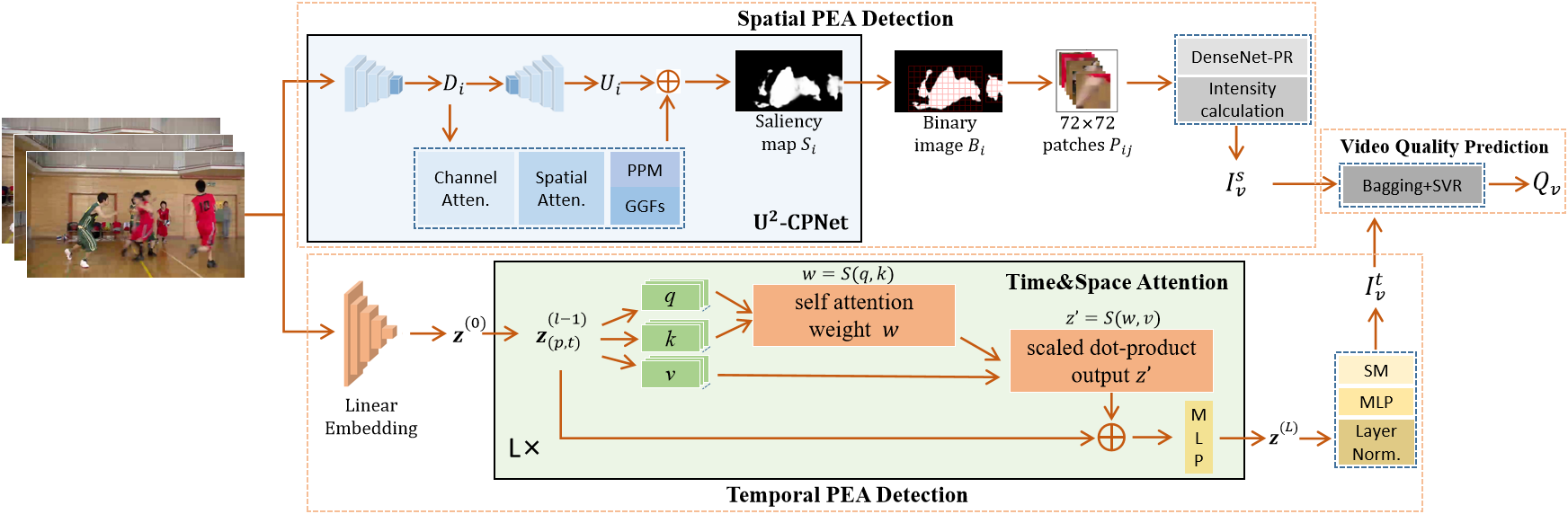}
	\setlength{\abovecaptionskip}{0.09cm}
	\caption{The overall framework of SSTAM. In spatial PEA detection, each video frame sequentially passes through the spatial PEA module to detect the intensities of spatial PEAs $I^{s}_{v}$. During the detection of temporal PEAs, $z^{(l-1)}$ in Transformer encoder block $l\in[1, L]$ passes through the temporal attention module, the spatial attention module, and the MLP in sequence. The output of temporal PEA module is $I^{t}_{v}$.}
	\label{fig2}
	\vspace{-0.2cm}
\end{figure*}

\noindent This section systematically introduces the framework of SSTAM, whose overall structure is depicted in Fig. \ref{fig2}. SSTAM consists of a spatial PEA detection model, a temporal PEA detection model and a video quality prediction model. 

\subsection{Spatial PEA Detection Model}
\noindent The spatial PEA detection model can be divided into visual saliency model and spatial PEA detection. First, the visual saliency model extracts the saliency areas. This model can exclude areas that have little impact on the visual experience, making it better suited to human visual perception while reducing computational complexity. Then, spatial PEAs within the saliency regions are obtained by the spatial PEA detection part.

\pmb{Visual Saliency Model:} An essential component of the Human Visual System (HVS) and a determinant of the perceived quality of videos is visual saliency. Saliency object detection identifies the most visually distinctive objects in an image or video. Focusing on saliency regions can also significantly reduce the computational burden on the network. Existing saliency models are focused on the extraction of a specific class of targets \cite{17, 18, 19, 20}. For videos, changes in target motion and appearance contrast are more attention-grabbing, so extracted saliency targets should be more oriented to such regions \cite{17}. Given the characteristics of compression artifacts, we focus more on channel attention, which is also known as feature attention. Channel linkage is used to reflect the importance of each feature. In addition, spatial attention is employed to complement channel attention. The channel and spatial features are extracted in this work using Convolutional Block Attention Module (CBAM) \cite{21}.

However, it is important to combine the extracted channel and spatial features with semantic information to enable the network to better identify saliency regions. Though high-level semantic features are helpful for locating saliency objects, semantic features at lower levels can also provide essential local details. The low-level semantic information is often diluted in the top-down process. The absence of local information often disturbs the localization of saliency objects. Drawing inspiration from U-Net \cite{22} and ${\rm U}^2$-Net \cite{23}, we extract multi-level features and fuse them together for better detection. For the information at each level to be used properly, Global Guiding Flows (GGFs) \cite{24} is utilized to determine the weights of the global guidance information for each part of the top-down path. It can integrate the information generated by the Pyramid Pooling Module (PPM) \cite{24} with the feature maps of the top-down pathway output. Inspired by this, we present an improved network ${\rm U}^2$-Convolutional Pyramid Network (${\rm U}^2$-CPNet), incorporating the attention mechanism and global guidance.

In addition, the loss function is also improved to make the model more target to identify the saliency regions. During the training process, the Binary Cross Entropy (BCE) loss is pixel-level, which does not consider the labels of adjacent regions and gives equal weights to the pixels in saliency regions and the background. This loss function aims to achieve convergence on all pixels, and does not fit well with the purpose of extracting saliency regions. Loss function should pay more attention to the correlation between adjacent pixels and the global aspect during training. In this paper, the BCE function is used to allow the network to fully acquire pixel-level data. The global background and the local information among neighboring pixels are addressed using intersection over union (IoU) and structural similarity (SSIM), respectively. IoU enables the network to attend more to the global context of the saliency map and concentrate on the foreground targets \cite{25}. The expression of BCE and IoU loss function is shown in Eq. \ref{eq_bce} and Eq. \ref{iou}.
\begin{equation}
	\label{eq_bce}
	L_{\rm bce} = -(1-G_{(a,b)})\log{(1-S_{(a,b)})}-G_{(a,b)}\log{S_{(a,b)}},
\end{equation}
\begin{equation}
	\label{iou}
	L_{\rm iou} = \frac{\sum\limits_{a=1}^{H} \sum\limits_{b=1}^{W} S_{(a,b)} G_{(a,b)}}{\sum\limits_{a=1}^{H} \sum\limits_{b=1}^{W} [S_{(a,b)}+G_{(a,b)}-S_{(a,b)}G_{(a,b)}]},
\end{equation}
where $a$ and $b$ represent the vertical and horizontal coordinates of the pixel in each video frame. $H$ and $W$ indicate the height and width of the video frame. $G_{(a,b)}\in\{0,1\}$ denotes the real label of pixel $(a,b)$ and $S_{(a,b)}\in\{0,1\}$ is the predicted probability of the saliency object.

SSIM deals with the structural relationships among adjacent pixels in the image content. We mix it into the training loss to highlight the structural information of the saliency objects. It ensures that there is enough space for feature exploitation. While the probability is reduced to zero in the detected regions of saliency, the backdrop is perceived visually as being considerably cleaner. The expression of the SSIM loss function is shown in Eq. \ref{ssim}.
\begin{equation}
	\label{ssim}
	L_{\rm ssim} = 1 - \frac{(2\mu_G \mu_P + C_1^2)(2\sigma_{GP} + C_2^2)}{(\mu_G^2 + \mu_P^2 +C_1^2)(\sigma_G^2 + \sigma_P^2 +C_2^2)},
\end{equation}
where the subscript $G$ denotes the binary ground truth mask, and the subscript $P$ is the predicted probabilistic map mask. The $\mu_G$ and $\mu_P$ refer to the mean value, $\sigma_G$ and $\sigma_P$ represent the standard deviation of both, and $\sigma_{GP}$ indicates as their covariance. $C_1$ and $C_2$ are 0.01 and 0.03 \cite{26}, respectively. They are employed to avoid the division by zero error.

The strategy for mixing losses during model training is shown in Eq. \ref{L}, where $L$ is the loss of the training process, and the weight of each component is 1.
\begin{equation}
	\label{L}
	L = L_{\rm bce} + L_{\rm iou} + L_{\rm ssim}.
\end{equation}

\pmb{Spatial PEA Detection:} As for spatial PEA recognition, DenseNet for PEA Recognition (DenseNet- PR) \cite{16} can alleviate the problem of gradient disappearance and improve feature transmission while greatly reducing the number of parameters. Video frames are pre-processed to fit the DenseNet-PR input: after obtaining the saliency map $S_i$ extracted by ${\rm U}^2$-CPNet, we binarize it into $B_i$, then map the $B_i$ back to the video frame and partition the saliency regions in the video frame into $72\times72$ patches. Four types of spatial PEAs recognition models are trained to detect the PEAs in patches. The probability list of each $72\times72$ patch, namely $I_{\rm ij}$, is obtained. From which the intensity of each spatial PEA in the whole video sequence is finally calculated through the following equation.
\begin{equation}
	\label{Ivs}
	\setlength{\abovedisplayskip}{3pt}
	\setlength{\belowdisplayskip}{3pt}
	I_v^s = \frac{1}{N_f N_p} \sum_{i=1}^{N_f} \sum_{j=1}^{N_p} I_{\rm ij},
\end{equation}
where $N_p$ refers to the total number of patches in the saliency regions of each frame. $N_f$ denotes the total number of video frames. $I_v^s$ represents the intensity value of spatial PEAs in each video. $I_{\rm ij}$, as mentioned above, denotes the probability of certain PEA of each $72\times72$ patch. $i$ is the $i$-th frame of the video and $j$ is the $j$-th patch of the video frame.

\subsection{Temporal PEA Detection Model}
\noindent

Video-specific tasks consider both spatial and temporal information. Compared to other video understanding tasks, temporal PEAs detection is more dependent on spatio-temporal features which are more challenging for the network to capture. Meanwhile, there is a great lack of research on the analysis and detection of temporal PEAs. Therefore, it is imperative to exploit temporal PEA detection. Self-attention mechanism can capture both local and global dependencies over a long range by directly comparing all spatio-temporal features. Hence, self-attention based TimeSFormer \cite{27} is improved and applied to the detection of temporal PEAs. Since the spatial features of temporal PEA are relatively insignificant, we focus more on temporal features when training the network.

Training data of flickering and floating is derived from the database PEA265 \cite{16}. Temporal PEA detection model is shown in Fig. \ref{fig2}. The intensity of the temporal PEA for each video is finally computed as follows.
\begin{equation}
	\label{Ivt}
	I_v^t = {\rm SoftMax}({\rm MLP}({\rm LN}(\sum\limits_{j=1}^{N_c} \sum\limits_{i=1}^{N_t} {\rm TSF}_j(F_i)))),
\end{equation}
where we divide the consecutive $N_t$ frames into a clip and $N_c$ is the total number of clips for each video. TSF() is the processing of TimeSFormer and LN() refers to Layer Norm. ${I}_{v}^{t}$ is the intensity of temporal PEA in each video.

\subsection{Video Quality Prediction}
\noindent

Predicting video quality requires effectively mapping the correlation between the spatial and temporal artifacts detected in the previous sections and the subjective quality scores. We use the Bootstrap Aggregating (Bagging) \cite{28} method in integrated learning. Multiple base learners using different parts of the training set are combined to achieve better generalization performance than a single learner. Assuming a complete dataset, $D$, we match the MOS$|$DMOS values to the intensities of the six PEAs of each video in $D$. The matching process is shown in the following equation.
\begin{equation}
	\label{D}
	D=\{(I_{V_{1}},{\rm MOS}_{1}|{\rm DMOS}_{1}),\ldots, (I_{V_{k}},{\rm MOS}_{m}|{\rm DMOS}_{m})\},
\end{equation}
where ${\rm MOS}_{m}|{\rm DMOS}_{m}$ denotes the subjective quality score of the $m$-th compressed video, $I_{V_k}$ is the summary of the intensities of all spatial and temporal PEAs.

For training, we divide the dataset into a training set and a testing set in a ratio of 8:2, followed by dividing the training set into ten sub-training sets evenly. Support Vector Regression (SVR) models are trained using these ten sub-training sets. The performance of these base learners is evaluated by the Pearson Linear Correlation Coefficient (PLCC) between the predicted and true quality scores. The top three learners in PLCC scores among the ten base learners are chosen to predict the video quality $Q_v$. The expression for $Q_v$ is shown below.
\begin{equation}
	\label{Qv}
	Q_v = \sum\limits_{i=1}^{n} \omega_iy_i(x),
\end{equation}
\begin{equation}
	\label{omega}
	\sum\limits_{i=1}^{n} \omega_i = 1 \quad \omega_i\in[0,1),
\end{equation}
where $x$ is testing set. $y_i(x)$ represents the prediction output of the $i$-th base learner. $\omega_i$ is the weight of the $i$-th base learner. $n$ denotes the number of base learners, which value is 10. 

\section{Experiments and Discussions}
Experiments and evaluations are carried out on four commonly used and publicly accessible Video Quality Databases (VQDs), including LIVE \cite{29}, CSIQ \cite{30}, IVP \cite{31}, and FERIT-RTRK \cite{32}, to verify the generalizability of SSTAM. LIVE contains 40 compressed videos with the resolution of $768\times432$. CSIQ consists of 36 compressed videos at a resolution of $832\times480$. IVP and FERIT-RTRK contain 40 and 30 compressed videos with the resolution of $1920\times1080$, respectively. In this work, we matched the six compression artifacts intensities of compressed videos from these four databases with their MOS$|$DMOS to form four complete datasets for evaluation. We compared SSTAM with several classical and popular quality assessment methods to verify its performance. These methods are PSNR, SSIM \cite{8}, MS-SSIM \cite{33}, STRRED \cite{9}, SpEED-QA \cite{10}, VIIDEO \cite{11}, TLVQM \cite{12}, SAAM \cite{14}, BRISQUE \cite{34} and NIQE \cite{35}. PLCC and Spearman Rank Correlation Coefficient (SRCC) values are employed to benchmark the performance of these methods.

\pmb{Performance Comparison:} The results can be seen in Tables \ref{tab1} and \ref{tab2}. The bolded result corresponds to the best performing method in each database. It is clear that our model is highly competitive on these four databases. Weighing the quantity of compressed videos in each database gets the overall performance. As can be seen from the last column, the total performance of SSTAM is better than the other methods. The high improvement of SAAM and SSTAM compared to other methods is because they fully consider the influence of PEA on compressed video quality. This experiment also fully demonstrates that there is a strong correlation between the PEAs and the subjective quality scores of compressed videos. We find that methods containing a saliency model predict video quality more effectively in FERIT-RTRK. Since the saliency objects in FERIT-RTRK are relatively few and large, the network can extract the saliency regions of the video more precisely. Accordingly, a better extraction of saliency regions contributes significantly to the performance of quality assessment.

\pmb{Ablation Experiments:} To further verify the improvement of our proposed strategies for video quality assessment, we compared the effects of saliency model and temporal PEAs separately. Visual saliency model called Attentive CNN-LSTM Network (ACLNet) \cite{36} was applied in SAAM. To evaluate the performance, we substitute ACLNet in SAAM with $\rm U^2$-CPNet. After changing the saliency network, we refer to the method as SAAM2.0. As shown in Table \ref{tab3}, our saliency model performs better in recognizing saliency regions and enhances the performance of the network. The comparisons of SAAM2.0 and SSTAM reveal that temporal PEAs have an effect. 
\begin{table}[!t]
	\caption{Performance comparison in terms of PLCC.\label{tab1}}
	\centering
	\vspace{-0.5em}
	\small\tabcolsep=0.13cm
	\begin{tabular}{ c c c c c c }
		\hline
		Methods & LIVE & CSIQ & IVP & FERIT-RTRK & Overall\\
		\hline
		PSNR & 0.5735 & 0.8220 & 0.7998 & 0.7756 & 0.7383\\
		SSIM & 0.6072 & 0.8454 & 0.8197 & 0.6870 & 0.7406\\
		MS-SSIM & 0.6855 & 0.8782 & 0.8282 & 0.8724 & 0.8105\\
		STRRED & 0.8392 & 0.8772 & 0.5947 & 0.8425 & 0.7823\\
		SpEED-QA & 0.7933 & 0.8554 & 0.6822 & 0.6978 & 0.7586\\
		BRISQUE & 0.2154 & 0.5526 & 0.2956 & 0.7653 & 0.4335\\
		NIQE & 0.3311 & 0.5350 & 0.3955 & 0.5817 & 0.4505\\
		VIIDEO & 0.6829 & 0.7211 & 0.4358 & 0.3933 & 0.5651\\
		TLVQM & 0.7511 & 0.7740 & - & - & 0.7619\\
		SAAM & 0.9023 & 0.9244 & 0.8717 & 0.9499 & 0.9091\\
		\hline
		SSTAM & \pmb{0.9490} & \pmb{0.9523} & \pmb{0.9478} & \pmb{0.9769} & \pmb{0.9552}\\
		\hline
		\vspace{-1.5em} 
	\end{tabular}
\end{table}
\begin{table}[!t]
	\caption{Performance comparison in terms of SRCC.\label{tab2}}
	\centering
	\vspace{-0.5em}
	\small\tabcolsep=0.13cm
	\begin{tabular}{ c c c c c c }
		\hline
		Methods & LIVE & CSIQ & IVP & FERIT-RTRK & Overall\\
		\hline
		PSNR & 0.4146 & 0.8028 & 0.8154 & 0.7685 & 0.6928\\
		SSIM & 0.5677 & 0.8440 & 0.8049 & 0.7236 & 0.7328\\
		MS-SSIM & 0.6773 & 0.9465 & 0.7917 & 0.8508 & 0.8107\\
		STRRED & 0.8358 & \pmb{0.9770} & 0.8595 & 0.8310 & 0.8761\\
		SpEED-QA & 0.7895 & 0.9639 & 0.8812 & 0.7945 & 0.8587\\
		BRISQUE & 0.2638 & 0.5655 & 0.1051 & 0.7574 & 0.3961\\
		NIQE & 0.1769 & 0.5012 & 0.2351 & 0.4855 & 0.3362\\
		VIIDEO & 0.6593 & 0.7153 & 0.1621 & 0.3177 & 0.4667\\
		TLVQM & 0.7338 & 0.7956 & - & - & 0.7631\\
		SAAM & 0.8691 & 0.8810 & 0.8413 & 0.9429 & 0.8796\\
		\hline
		SSTAM & \pmb{0.9201} & 0.9177 & \pmb{0.9048} & \pmb{0.9823} & \pmb{0.9281}\\
		\hline
		\vspace{-1.5em} 
	\end{tabular}
\end{table}
\begin{table}[!t]
	\caption{ablation experiments.\label{tab3}}
	\centering
	\vspace{-0.5em}
	\small
	\begin{tabular}{ c c c c }
		\hline
		Indicators & SAAM & SAAM2.0 & SSTAM\\
		\hline
		PLCC & 0.9091 & 0.9475 & 0.9552\\
		SRCC & 0.8796 & 0.9109 & 0.9281\\
		\hline
		\vspace{-1.5em}
	\end{tabular}
\end{table}

\section{Conclusions}
\noindent This paper proposes a compressed video quality index based on saliency-aware spatio-temporal artifact detection. The experimental results demonstrate the improved performance and generalizability of the proposed SSTAM. We believe it will be essential in compressed video quality assessment and video coding optimization in the future.

\end{document}